\documentclass[9pt,twocolumn,twoside]{pnas-new}
\usepackage{import}
\templatetype{pnasresearcharticle} 

\title{Time-Resolved Two Million Year Old Supernova Activity Discovered in the Earth’s Microfossil Record}

\author[a]{Peter Ludwig}
\author[b,1]{Shawn Bishop} 
\author[b]{Ramon Egli}
\author[a]{Valentyna Chernenko}
\author[a]{Boyana Deneva}
\author[a]{Thomas Faestermann}
\author[a]{Nicolai Famulok}
\author[a]{Leticia Fimiani}
\author[a]{José Manuel Gómez-Guzmán}
\author[a]{Karin Hain}
\author[a]{Gunther Korschinek}
\author[c]{Marianne Hanzlik}
\author[d]{Silke Merchel}
\author[d]{Georg Rugel}

\affil[a]{Physik Department, Technische Universität München, 85748 Garching, Germany}
\affil[b]{Geomagnetism and Gravimetry, Central Institute for Meteorology and Geodynamics, 1190 Vienna, Austria}
\affil[c]{Chemie Department, FG Elektronenmikroskopie, Technische Universität München, 85748 Garching, Germany}
\affil[d]{Helmholtz-Zentrum Dresden-Rossendorf, Helmholtz Institute Freiberg for Resource Technology, 01328 Dresden, Germany}

\leadauthor{Ludwig} 

\significancestatement{Massive stars, which terminate their evolution in a cataclysmic explosion called a type-II supernova, are the nuclear engines of galactic nucleosynthesis.  Among the elemental species known to be produced in these stars, the radioisotope $^{60}$Fe stands out: this radioisotope has no natural, terrestrial production mechanisms and, thus, a detection of $^{60}$Fe atoms within terrestrial reservoirs is proof for the direct deposition of supernova material within our Solar System. We report in this work the direct detection of live $^{60}$Fe atoms in biologically-produced nano-crystals of magnetite, which we selectively extracted from two Pacific Ocean sediment cores. We find that the arrival of supernova material on Earth coincides with the lower Pleistocene boundary (2.7~Ma) and that it terminates around $1.7$~Ma.}

\authorcontributions{P.L. performed the magnetic measurements, prepared the $^{60}$Fe AMS samples, analysed the AMS data, participated in the AMS measurements and co-wrote the manuscript. S.B. supervised the overall project and co-wrote the manuscript; R.E. performed the analysis of the magnetic measurements; and co-wrote the manuscript; V.C. prepared $^{60}$Fe AMS samples; B.D., T.F., N.F., L.F., J.M.G-G, K.H, and G.K made essential contributions to the AMS measurements and operation of the MLL AMS facility; M.H. performed the electron microscopy work and imaging; S.M. and G.R. helped to develop the CBD protocol and performed the $^{10}$Be/$^9$Be measurements. S.M. performed the $^{10}$Be sample preparation. All authors contributed to the discussion about the final manuscript and its contents.}

\authordeclaration{The authors declare no conflict of interest.}
\correspondingauthor{\textsuperscript{1}To whom correspondence should be addressed. E-mail: shawn.bishop@ph.tum.de}

\keywords{accelerator mass spectrometry $|$ magnetofossils $|$ supernova} 

\begin{abstract}
Massive stars ($M\gtrsim10~M_\odot$), which terminate their evolution as core collapse supernovae, are theoretically predicted to eject $>10^{-5} M_\odot$ of the radioisotope $^{60}$Fe (half-life $2.61$~Ma). If such an event occurs sufficiently close to our solar system, traces of the supernova debris could be deposited on Earth. Herein, we report a time-resolved $^{60}$Fe signal residing, at least partially, in a biogenic reservoir. Using accelerator mass spectrometry, this signal was found through the direct detection of live $^{60}$Fe atoms contained within secondary iron-oxides, among which are magnetofossils; the fossilized chains of magnetite crystals produced by magnetotactic bacteria. The magnetofossils were chemically extracted from two Pacific Ocean sediment drill cores. Our results show that the $^{60}$Fe signal onset occurs around $2.6-2.8$~Ma, near the lower Pleistocene boundary, terminates around $1.7$~Ma, and peaks at about $2.2$~Ma.
\end{abstract}

\dates{This manuscript was compiled on \today}
\doi{\url{www.pnas.org/cgi/doi/10.1073/pnas.1601040113}}

\begin{document}

\verticaladjustment{-2pt}

\maketitle
\thispagestyle{firststyle}
\ifthenelse{\boolean{shortarticle}}{\ifthenelse{\boolean{singlecolumn}}{\abscontentformatted}{\abscontent}}{}

\dropcap{T}he isotope $^{60}$Fe is mostly produced during the evolution of massive stars ($M\gtrsim10~M_\odot$) in two steps along the course of their evolution. In the first step, during their quiescent helium and carbon shell burning phases, $^{60}$Fe is synthesized at the base of these shells by a slow neutron capture process on pre-existing stable seed nuclei, mostly by two successive neutron captures starting on stable $^{58}$Fe \citep{Lim06}. In the second step, shock heating within the carbon-shell, driven by the passage of the supernova (SN) blast wave, allows for the synthesis of a modest amount of $^{60}$Fe just prior to the subsequent disruption of the star and the concomitant explosive ejection of these shell mass zones, and the $^{60}$Fe within them, into the interstellar medium. The subsequent beta-decay of $^{60}$Fe, which has a half-life of $t_{1/2}=(2.61 \pm 0.04)~$Ma \cite[weighted average of refs.][]{Rug09,Wal15b}, gives rise to two characteristic gamma-ray lines, which have been detected in the central part of the galactic plane \cite{Wan07}, known to be a site of massive stars, confirming the association of $^{60}$Fe formation with regions of ongoing massive star nucleosynthesis.

If a core-collapse supernova (CCSN) occurs  sufficiently close to our solar system, part of the ejected matter should arrive in our solar system. The best candidate mechanism for overcoming the solar wind pressure and penetrating to the Earth’s orbit is dust transport \citep{Kni99,Ath11}. Recent investigations with far-infrared to sub-millimeter telescopes suggest that CCSN ejecta are efficient dust sources \citep{Gom14}. For instance, copious amounts of dust with a large mass fraction contained in grains of above $0.1~\mu $m size (and up to $4.2~\mu$m) have recently been observed in the ejecta of supernova 2010jl \citep{Gal14}. During atmospheric entry, dust grains are expected to be partially or totally ablated, depending on composition, incident velocity, and angle of entry \citep{Pla12}. The $^{60}$Fe released in the ablated fraction will enter the terrestrial iron cycle \citep{Jic05} and become deposited into geological reservoirs such as marine sediments. 

An excess of $^{60}$Fe was already observed in $\sim 2$~Ma old layers of a ferromanganese (FeMn) crust retrieved from the Pacific Ocean \cite{Kni99,Kni04,Fit08} and recently in lunar samples \cite{Fim16}. However, due to the slow growth rate of the FeMn crust, the $^{60}$Fe signal had a poor temporal resolution. This $^{60}$Fe has been attributed to a deposition of SN ejecta; though, this interpretation has been challenged by an alternate hypothesis attributing the $^{60}$Fe excess to micrometeorites \cite[][and references therein]{Stu12}. An independent indication of a recent SN interaction with our solar system was recently deduced from the spectra of cosmic ray particles \cite{Kac15}. In our work herein, we aimed at the analysis of the entire temporal structure of the $^{60}$Fe signature in terrestrial samples. This requires a geological reservoir with an excellent stratigraphic resolution, high $^{60}$Fe sequestration and low Fe mobility, which preserves $^{60}$Fe fluxes as they were (or nearly so) at the time of deposition, apart from radioactive decay. Both conditions were fulfilled in the carefully selected marine sediments used in this study.

One particularly interesting mechanism by which Fe is sequestered within sediments is through biomineralization, e.g. by dissimilatory metal reducing bacteria (DMRB) \cite{Zac02}, and by magnetotactic bacteria (MTB) \cite{Bel09a}. MTB are single-cell prokaryotes, which produce intracellular chains of magnetite (Fe$_3$O$_4$) nanocrystals called magnetosomes \cite{Bla75,Fai08}. These bacteria achieve their highest population densities near the so-called oxic-anoxic transition zone  \cite{Baz04}, where the oxygen concentration drops, producing a well-defined redox boundary. In pelagic sediments, this boundary occurs within few centimeters below the sediment-water interface \cite{Pet93}, forcing the MTB populations to move upwards as the sediment column grows, with dead cells being left behind. After decomposition of the dead cells, the magnetosomes remain embedded within the sediment bulk and are subsequently called magnetofossils \cite{Kop08}.

The ablated Fe fraction of SN dust grains arriving in the oceans is expected to undergo dissolution and re-precipitation upon reaching the sediment in the form of nano-minerals, such as poorly crystalline ferric hydroxides \cite{Jic05}. The poor solubility of Fe(III) minerals at circumneutral pH values ($\sim 0.1$~nmol/L) means that Fe is hardly mobilizable under oxygenated conditions. Many microorganisms, including DMRB and MTB, get around this problem by excreting organic compounds known as siderophores, which specifically complex Fe(III). DMRB reduce Fe(III) to highly soluble Fe(II), which in turn leads to the precipitation of new minerals, among which is magnetite (Fe$_3$O$_4$) \cite{Zac02}. Other bacteria perform Fe(II) oxidation \cite{Kon11}. The Fe uptake capability of DMRB from particulate sources depends on the type of source mineral and particle size. Poorly crystalline hydroxides, such as ferrihydrite (FeOOH), are preferred over goethite ($\alpha$-FeOOH) and hematite ($\alpha$-Fe$_2$O$_3$), and Fe reduction rates are proportional to the specific surface area of particles \cite{Rod96,Zac98,Bos10}. In particular, surface normalized bacterial Fe-reducing reaction rates are $\sim 1.5-2$ orders faster \cite{Bos10} for nano-sized ferrihydrite particles as compared to grains with sizes comparable to bulk detrital grains. The Fe uptake capability of MTB has been investigated less extensively. Common constituents of the DMRB iron metabolism, such as genes for ferrous and ferric iron uptake, siderophore synthesis, iron reductases, as well as iron-regulatory elements, are present in MTB \cite{Fai08}. MTB are therefore able to take up ferric and ferrous iron from various sources \cite{Yan12} with similar capabilities as for DMRB. This is also supported by the fact that MTB are the main source of one reduction product – magnetite – in many types of sediments \cite{Rob12,Lud13,Egl04,Egl10}. The combination of Fe reducing and oxidizing reactions supports the Fe cycle in sediments, yielding ultrafine ($<100$~nm) secondary Fe minerals \cite{Sob02, Tay11} (see Supporting Information).

MTB extant at the time of supernova $^{60}$Fe input to the ocean floor are expected, therefore, to have incorporated $^{60}$Fe into their magnetosomes, biogenically recording the supernova signal. Unlike other Fe minerals and biomineralization products, magnetofossils have unique magnetic signatures enabling their detection down to mass concentrations in sediment of the order of few ppm \cite{Lud13}. As an essential point for this work, magnetofossil preservation ensures that the $^{60}$Fe signal is not altered, because the very existence of these microfossils means that post-depositional Fe mobilization through reductive diagenesis \cite{Row09,Rob12} did not occur. 

In general, Fe(III) sources accessible to bacterial reduction can be extracted with buffered solutions of reducing agents such as citrate-bicarbonate-dithionite (CBD) and, to a lesser extent, ammonium oxalate \cite{Rod04}. The CBD protocol \cite{Meh58} has been specially conceived to selectively dissolve fine-grained ($<200~$nm) secondary oxides in soils, but not larger particles of lithogenic origin \cite{Hun95,Vid00} (see Supporting Information). Therefore, CBD can be used to selectively extract Fe from magnetofossils and other secondary minerals along with $^{60}$Fe, thereby minimizing any dilution from large-grained, $^{60}$Fe-free primary mineral phases.

\section*{Materials and Methods}

\begin{figure}
\centering
\includegraphics[width=1\linewidth]{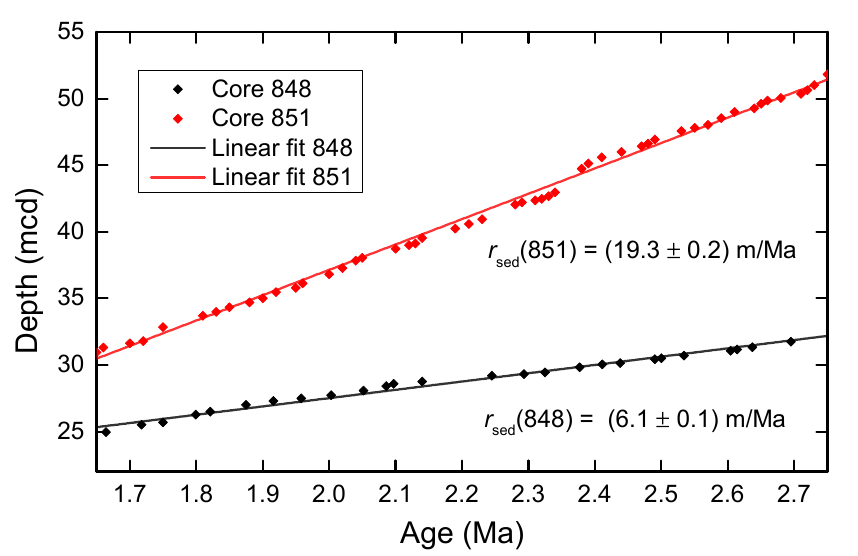}
\caption{Age model data \cite[taken from][]{Sha95} of ODP drill cores 848 and 851 with linear fits overlaid. The X-axis represents the depth in the respective drill cores in meters core depth (mcd). The slopes of the fits yield sedimentation rates as shown. No significant departure from linearity is observed across the displayed time range, indicating both cores had constant sedimentation rates. The errors represent 1-$\sigma$ standard deviations.}
\label{fig:Fig1}
\end{figure}

In our search for a biogenic supernova signal, we selected two sediment cores, core 848 and core 851, from the equatorial Pacific (2\textdegree59.6’S, 110\textdegree29’W, 3.87 km water depth), recovered by the Ocean Drilling Program during Leg 138 \cite{May92}. The sediment of both cores is a pelagic carbonate (60–80\% CaCO$_3$, 20–30\% SiO$_2$) with a total iron content of 1.5–3.5 wt\% \cite{Bil95}. Both cores are characterized by an excellent bio- and magnetostratigraphic record and almost constant sedimentation rates of ($6.1~\pm~0.1$)~m/Ma and ($19.3~\pm~0.2$)~m/Ma for core 848 and 851, respectively, over the 1.7-2.7~Ma age range of interest \cite{Sha95}, as shown in Fig.~\ref{fig:Fig1}.

The presence of magnetofossils in these cores was confirmed by electron microscopy and magnetic analysis techniques based on first-order reversal-curve measurements \cite{Bis11,Lud13} (see Supporting Information). The average magnetofossil Fe concentration over the 1.7–3.4~Ma interval was determined to be 25–30~$\mu$g/g for core 848, and 15–20~$\mu$g/g for core 851, respectively. Along with a constant sedimentation rate (Fig.\ref{fig:Fig1}) and sediment composition~\cite{Farrell95}, the lack of significant magnetofossil concentration variations, both absolute and relative to other magnetic Fe minerals, indicates that the depositional environment was stable during the period of time period under investigation. An additional confirmation of stable sedimentation conditions was obtained by measurements of the $^{10}$Be/$^{9}$Be ratio in representative samples of core 851, which show no significant deviation from exponential radioactive decay (see Supporting Information).


In order to maximize the $^{60}$Fe/Fe atom ratio, a highly selective version of the CBD procedure \cite{Meh58} was developed (see Supporting Information). This is a very mild chemical leaching technique, able to completely dissolve secondary iron oxides such as magnetofossils, while leaving larger grains, which may not contain $^{60}$Fe, essentially intact \cite{Lud13}. At least 27\% of the total Fe extracted with this technique is contained in magnetofossils \cite{Lud13}, which therefore represent a significant contribution to the analyzed Fe pool. Even with this carefully designed extraction protocol, the expected $^{60}$Fe concentrations are so low that the only viable measurement technique is ultra-sensitive AMS, which has been carried out at the GAMS (Gas-filled Analyzing Magnet System) setup \cite{Kni00} at the Maier-Leibnitz-Laboratory (MLL) in Garching (Germany) over several beamtimes. The MLL features a 14~MV MP Tandem accelerator with an energy stability of $\Delta E/E\approx 10^{-4}$. For AMS measurements of $^{60}$Fe, the samples are prepared as Fe$_2$O$_3$ powder mixed 50/50 by volume with Ag powder (-120 mesh, Alfa Aesar, Lot Nr.~J07W011), which is subsequently hammered into a 1.5~mm wide hole, drilled into a silver sample holder.

Fe was extracted as FeO$^-$ from a single-cathode, cesium sputter ion source, and injected into the accelerator after a pre-acceleration to 178~keV. The tandem terminal voltage for the experiments described here was between 11~MV and 12~MV, which favored selection of the charge state 10$^+$ for the radioisotope $^{60}$Fe after passing through a 4~$\mu$m thick carbon stripper foil at the accelerator terminal. A charge state of 9$^+$ was selected for the stable beam of $^{54}$Fe, since it has nearly the same magnetic rigidity as $^{60}$Fe$^{10+}$ at the same terminal voltage. After acceleration to 110–130~MeV (depending on the available terminal voltage), the ions exit the accelerator and pass another dipole magnet for selection of the correct magnetic rigidity, as well as two Wien-velocity filters. Finally, the ions are directed towards the dedicated GAMS (Gas-filled Analyzing Magnet System) beamline by way of a switching magnet. Through use of the GAMS magnet, the challenging suppression of the stable isobar $^{60}$Ni is achieved.

The GAMS magnet chamber is filled with  4–7~mbar of N$_2$ gas. Through electron exchange reactions with the N$_2$ gas, each ion species adopts an equilibrium charge state which depends on its atomic number $Z$. Thus, isobars will be forced on different trajectories if a magnetic field is applied. Since $^{60}$Fe and $^{60}$Ni have $\Delta Z=2$, their trajectories on the exit-side of the magnet can be spatially separated by 10~cm. The GAMS magnetic field can then be adjusted to make $^{60}$Fe to reach the particle identification detector, while most $^{60}$Ni is blocked using a suitable aperture in front of the detector entrance. The detector itself is an ionization chamber featuring a Frisch-grid and a five-fold split anode and is filled with $30 - 50$~mbar isobutane as counting gas. Individual $^{60}$Fe ions can thus be identified by their energy deposition (Bragg-curve), which is described in more detail in the Supporting Information.

For measurements of the atom ratio $^{60}$Fe/Fe, the stable beam of $^{54}$Fe$^{9+}$ is tuned into a Faraday cup in front of the GAMS, where the typical current is $30-150~$enA (sample dependent). Then, the terminal voltage and the injector magnet are switched to allow $^{60}$Fe$^{10+}$ to pass and the Faraday cup is retracted. $^{60}$Fe ions are then individually identified and counted in the ionization chamber, while the $^{60}$Ni background count-rate is only about $10-100~$Hz. In this manner, a highly selective discrimination of $^{60}$Fe against $^{60}$Ni and other background sources is achieved, as demonstrated by the extremely low blank levels obtained with different blank materials, such as processing blanks and environmental samples. The concentration of $^{60}$Fe/Fe is calculated from the number of $^{60}$Fe events counted by the detector, the measurement time, and the average current of $^{54}$Fe$^{9+}$ in front of the GAMS. The transmission between the Faraday cup and the detector is canceled out by relating the result to the known concentration of a standard sample, which is measured periodically during a beamtime. For this work, the standard sample PSI-12, with a concentration $^{60}\text{Fe/Fe}=(1.25 \pm 0.06)\times10^{-12}$, was used (see Supporting Information). The transmission efficiency of the entire system (including ion source yield, stripping yield, ion-optical transmissions, and software cuts) during $^{60}$Fe measurements is in the range $(1-4)\times 10^{-4}$.

\section*{Results}

\begin{figure}[t!]
\centering
\includegraphics[width=1\linewidth]{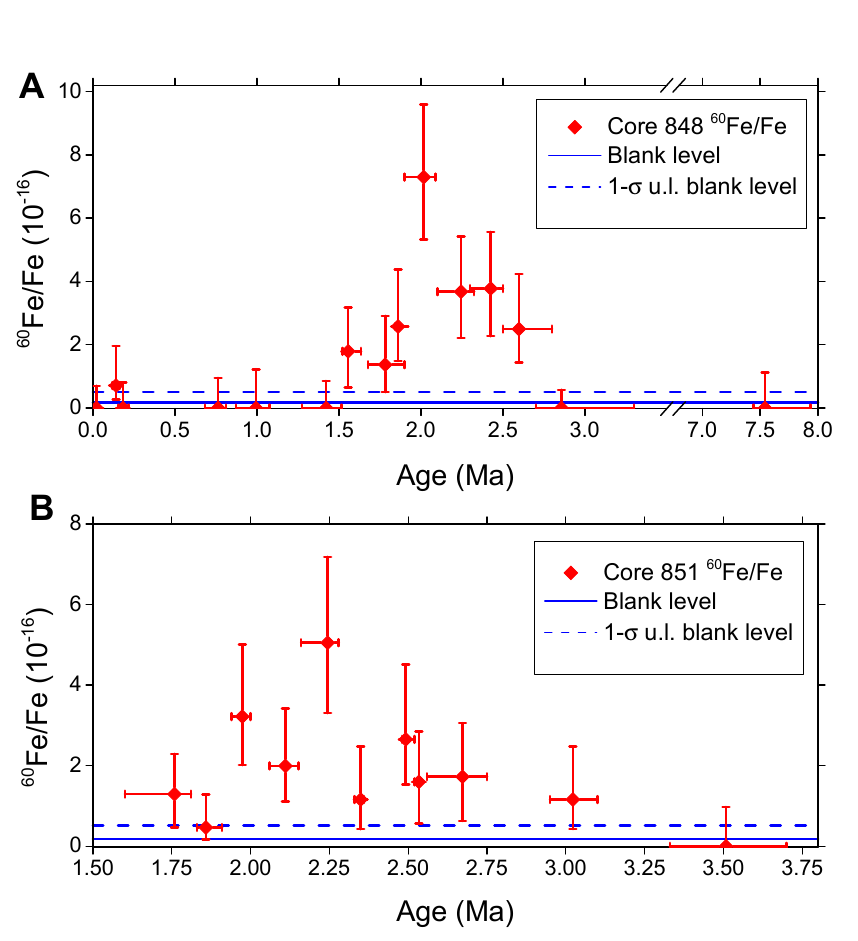}
\caption{$^{60}$Fe/Fe atom ratio determined in all sediment samples of core 848 ({\bf{A}}) and 851 ({\bf{B}}). Y-error bars indicate 1-$\sigma$ statistical uncertainties. The X-axes limits were chosen according to the availability of sample material, which also explains the broke X-axis in ({\bf{A}}). X-error bars represent core depth of sample material used for each data point. Each data point contains 3-6 adjacent individual AMS samples grouped together. The blank level and its 1-$\sigma$ upper limit (u.l.) are shown for a chemical processing blank. Note that panel  (\textbf{A}) has a broken time axis relative to panel (\textbf{B}).}
\label{fig:Fig2}
\end{figure}

A total of 111 sediment samples (67 from core 848 and 44 from core 851), each with a mass of $\sim 35~$g, were treated with the CBD protocol, yielding $\sim 5$~mg AMS samples consisting of Fe$_2$O$_3$. Each AMS sample was measured for an average of 4 hours until the sample material was exhausted, yielding one $^{60}$Fe event on average and a total of 86 events integrated over both cores (42 in core 848, 47 in core 851). Thus, several AMS samples have been grouped together to increase counting statistics, as displayed in Fig.~\ref{fig:Fig2}. Owing to the availability of several near-surface ($0-1$~Ma) and very deep ($7-8$~Ma) samples in core 848, the presence of a distinct $^{60}$Fe signal could be clearly identified (Fig.~\ref{fig:Fig2}A). The data is complemented by the observation of a similar signal in core 851 (Fig.~\ref{fig:Fig2}B), which is characterized by a $\sim$~1.5 times lower $^{60}$Fe/Fe ratio. The onset of the $^{60}$Fe signal occurs at $(2.7\pm 0.1)$~Ma and is centered at ($2.2\pm0.1$)~Ma. The signal termination is not as clear, since it remains slightly above the 1-$\sigma$ blank level until around 1.5~Ma, according to the data grouping used in Fig.~\ref{fig:Fig2}A. A detailed analysis averaging over both sediment cores and several data groupings yields a more conservative estimate for the termination time of ($1.7\pm 0.2$)~Ma. This results in a ($1.0\pm0.3$)~Ma long exposure of the Earth to the influx of $^{60}$Fe. 

An overview of the collected AMS data split into three age regions ($<1.8$~Ma, $1.8-2.6$~Ma, $> 2.6$~Ma) is shown in Tab.~\ref{Tab:1}. An estimate for the significance of the $^{60}$Fe signal in the peak region can be obtained by applying the procedure suggested by ref.~\cite{Cou08} for the superposition of two Poisson processes (i.e. signal and background). As a control region (no expected signal), we selected (1) the processing blank and, for comparison, (2) the $< 1.8$~Ma age interval of each respective sediment core. The second choice is rather conservative, since this overestimates the background signal. In the case of (1), the procedure yields a significance (in multiples of $\sigma$) of 5.2 and 3.9 (for core 848 and 851, respectively). In the case of (2) these values become 7.2 and 2.0. The significance for core 851 is low in the case of (2), since only little data was collected in the control region. 
 
A useful measure for the intensity of the total $^{60}$Fe exposure
is given by the terrestrial ($\Phi_{\text{ter}}$) fluence of $^{60}$Fe. $\Phi_{\text{ter}}$ represents the time-integrated, decay corrected flux of $^{60}$Fe into a given terrestrial reservoir over the entire exposure time. In order to sum up the contributions of all sediment layers in the signal range, the following integral is computed to calculate the average concentration of $^{60}$Fe/Fe in the signal range:

 \begin{equation}
 \overline{\mathcal{C}}=(t_2-t_1)^{-1} \int_{t_1}^{t_2} \mathcal{C} (^{60}\text{Fe/Fe}) \text{d}t
 \end{equation}
 

\noindent The average terrestrial $^{60}$Fe fluence, $\Phi_{\text{ter}}$, is then
given by

\begin{equation}
\Phi_{\text{ter}}=\overline{\mathcal{C}}\rho r_\text{sed}(t_2-t_1) Y_{\text{CBD}} N_{\text{A}}/W_{\text{Fe}}
\end{equation}

\noindent where $\rho$ is the dry sediment density, $r_\text{sed}$ is the sedimentation rate, $Y_{\text{CBD}}$ is the efficiency corrected CBD yield of extracted iron from per unit mass of dry sediment, $N_{\text{A}}$ is Avogadro’s number, and $W_{\text{Fe}}$ is the molecular weight of iron. Using $(t_2-t_1)=(1.0\pm0.3)$~Ma as a nominal signal duration results in a terrestrial fluence into our sediments of $\Phi_{\text{ter}}^\text{848}=(4.7\pm 1.6)\times10^5$~at/cm$^2$ and $\Phi_{\text{ter}}^\text{851}=(8.8\pm 2.9)\times10^5$~at/cm$^2$ for cores 848 and 851, respectively. The slightly different values and larger errors compared to Tab.~\ref{tab:Tab1} result from an averaging over different exposure times $(t_2-t_1)$, whereas the fluences in Tab.~\ref{tab:Tab1} were calculated for a fixed $(t_2-t_1)=0.8~$Ma. For the following discussion, we use the error-weighted mean of the fluences determined in both sediment cores, which is $\Phi_{\text{ter}}^\text{sed}=(5.6\pm 1.4)\times10^5$~at/cm$^2$.

In order to compare this result with the fluence obtained by ref.~\cite{Kni04} for the Pacific Ocean FeMn crust, several correction factors must first be applied. Since those results were published, the half-life of $^{60}$Fe has undergone a revision \cite{Rug09,Wal15b}; the half-life of $^{10}$Be, which was used to convert the crust growth rate to geological time, has also been re-determined \cite{Chm10,Kor10}. A correction for the different $^{60}$Fe standard samples used (which became available due to advanced cross-calibration measurements), must be taken into account. The resulting, decay corrected, terrestrial fluence derived from the FeMn crust now becomes $\Phi_{\text{ter}}^\text{crust}=(2.5\pm1.3)\times10^6$~at/cm$^2$, which is about a factor of $4-5$ higher than our result and does not take an uptake efficiency for Fe of the FeMn crust into account. Interestingly, another recently reported fluence value deduced from Indian Ocean sediments \cite{Wal16} ($\Phi_{\text{ter}}^\text{ind}=(35.4\pm 2.6)\times10^6$~at/cm$^2$) is 1-2 orders of magnitude higher than the value reported in this work. Possible explanations for such differences are (1) a non-uniform $^{60}$Fe deposition at the respective locations of the geological reservoirs \cite[see e.g.][]{Fry16} and (2) a selective chemical Fe uptake from bottom water currents at certain locations leading to fluence values above those related to the depositional flux.

\begin{table*}
\centering
\caption{Summary of all $^{60}$Fe data presented in this work. Lines indicate separation of the sediment cores into three intervals. The last line represents data obtained from processing blank material. The corrected (corr.) $^{60}$Fe/Fe concentrations are corrected for radioactive decay and the blank level.  All errors are given as 1-$\sigma$ uncertainties.}
\begin{tabular}{lrrrrrrr}
Sample & Age range & Number of & $^{60}$Fe events & $^{60}$Fe/Fe AMS & $^{60}$Fe/Fe corr. & Deposition rate & Terrestrial fluence $\Phi_\text{ter}$  \\
& (Ma) & samples & & ($\times10^{-16}$) & ($\times10^{-16}$) & (at. $^{60}$Fe a$^{-1}$~cm$^{-2}$) & ($10^5$ at.$^{60}$Fe cm$^{-2}$)  \\
\midrule
Core 848 & $0.0-1.8$ & $25$ & $4$ & $0.3^{+0.2}_{-0.1}$ & $0.2^{+0.3}_{-0.2}$ & $0.02^{+0.03}_{-0.02}$ & $0.10^{+0.18}_{-0.10}$ \\
Core 848 & $1.8-2.6$ & $27$ & $38$ & $4.3\pm 0.9$ & $7.4 \pm 1.2$ & $0.50 \pm 0.14$ & $4.0 \pm 1.0$ \\
Core 848 & $2.6-8.0$ & $15$ & $0$ & $<0.3$ & $<0.9$ & $<0.06$ & $<0.48$ \\
\midrule
Core 851 & $1.6-1.8$ & $3$ & $2$ & $1.6^{+1.8}_{-1.0}$ & $2.4^{+2.9}_{-1.6}$ & $0.47^{+0.56}_{-0.32}$ & $3.4^{+4.5}_{-2.5}$\\
Core 851 & $1.8-2.6$ & $22$ & $40$  & $2.7\pm 0.4$ & $4.6\pm 0.8$ & $0.92 \pm 0.23$ & $7.5 \pm 2.5$  \\
Core 851 & $2.6-3.7$ & $19$ & $5$  & $1.1^{+0.6}_{-0.5}$ & $2.2^{+1.4}_{-1.1}$ & $0.44^{+0.26}_{-0.22}$ & $3.5^{+2.0}_{-1.8}$\\
\midrule
Blank & & $18$ & $1$  & $0.2^{+0.3}_{-0.1}$ & &\\
\bottomrule
\label{Tab:1}
\end{tabular}

\label{tab:Tab1}
\end{table*}

\section*{Conclusions}
In summary, we have extracted magnetofossils and other iron-bearing secondary minerals from two Pacific Ocean sediment cores, 848 and 851 of the ODP Leg 138. Using AMS on samples derived from these mineral phases, we have detected a time-resolved $^{60}$Fe signal; and this signal coincides with independently observed ones in a deep ocean FeMn crust \cite{Kni04,Fit08}. In view of our results here, we would also note that the North Atlantic sediment results of ref. \cite{Fit08} possibly show a weak $^{60}$Fe signal in the same time range as ours. More recently, two other studies have found a compatible $^{60}$Fe signature in lunar samples \cite{Fim16}; and in Indian Ocean marine sediments \cite{Wal16}, with high statistical significance. Our results derive  from two independent Pacific Ocean sediment cores, each possessing a continuous time record spanning the entire time window of the $^{60}$Fe signal and with no detectable changes in sedimentation rate (Fig.~\ref{fig:Fig1},and Fig.~S8), no detectable changes of the depositional environment, and a relatively constant concentration of magnetofossils, respectively, (see Fig.~S1 and ref.~\cite{Bis11}) across the time span of the $^{60}$Fe signal. For these reasons, our cores 848 and 851 data should be faithful recorders of the temporal profile of the SN material during its arrival on Earth. Because our Fe extraction protocol specifically targets authigenic Fe oxides including magnetofossils, rather than the total Fe mineral pool (which is $\approx 3\%$ of dry sediment mass), as would aggressive leaching procedures, we potentially avoided a $^{60}$Fe dilution by up to $\sim 2$ orders of magnitude, because the majority of the total Fe mass is of detrital origin. This conclusion is supported by the fact that only $\approx 1\%$ of the total Fe mineral pool is extracted by the CBD procedure (see Supporting Information), which, as seen with magnetic minerals, does not leach lithogenic minerals. Such a strong dilution would have placed our $^{60}$Fe/Fe ratio below the blank level, and thereby beyond detectability.


We attribute this $^{60}$Fe signal to SN provenance, rather than to micrometeorites, for the following reasons: first, MTB are expected to obtain their iron budget from poorly crystalline hydroxides (see Supporting Information), and not from silicate and magnetite micrometeorite grains of $>~20~\mu$m diameter \cite{Stu12}; second, our CBD protocol was designed to selectively dissolve only fine grain magnetite $<200$~nm in size, thereby avoiding dissolution of any large-scale micrometeorites. Thus, the $^{60}\mbox{Fe}$ we have extracted cannot be from such micrometeorites.
 
The Local Bubble~\cite{appel01} is a low density cavity $\sim 150$~pc in diameter, within the interstellar medium of our galactic arm, in which the solar system presently finds itself. It has been carved out by a succession of $\sim 20$ supernovae over the course of the last $\sim 10$~Ma likely having originated from progenitors in the Scorpius-Centaurus OB star association~\cite{appel01,Ben02}; a gravitationally unbound cluster of stars $\sim 50$~pc in radius. Analyses~\cite{Breitschwerdt16,Ben02} of the relative motion of this star association has shown that around 2.3~Ma, it was located at minimum distance, of $\sim 100$~pc from the solar system, making it the most plausible host for any supernova responsible for the  $^{60}\mbox{Fe}$ signal. The geological time span covered by our  $^{60}\mbox{Fe}$ signal is intriguing, in that there is an established and overlapping marine extinction event of mollusks~\cite{Allmon93,Pet95}, marine snails~\cite{Dietl04} and bivalve fauna~\cite{Raffi85,Stanley86}, in addition to a coeval global cooling period~\cite{Berkman96,Raffi85,Stanley86}. The question whether this supernova could have contributed to this extinction has been previously raised~\cite{Ben02}, where it is considered that a supernova-induced UV-B catastrophe~\cite{Cockell99}, and its concomitant knock-on effects in the marine biosphere via phytoplankton die off, could have been a proximate factor in this extinction.  However, consensus now indicates that a canonical CCSN would need to be within 10~pc of our Solar System for there to be a significant and lasting depletion of the ozone layer strong enough to give rise to sudden extinction events~\cite{Geh03,Beech11,Mellott16}, disfavoring a direct and sudden causal connection, such as a UV-B catastrophe, between a Scorpius-Centaurus CCSN and the Plio-Pleistocene marine extinction.

\acknow{The search for $^{60}$Fe supernova signatures is supported by the German Research Foundation (DFG), Grant DFG-Bi1492/1-1 and by the DFG Cluster of Excellence “Origin and Structure of the Universe” (www.universe-cluster.de). We are grateful to three anonymous reviewers for their constructive comments, which helped improve the original manuscript and to the operators of the MLL and the Ion Beam Center at HZDR for their help during beamtimes.}

\showacknow 


\nocite{*}
\bibliography{PNAS-template-main}

\verticaladjustment{-2pt}


\section{Sedimentary Fe cycle and Fe-reducing bacteria}

\subsection*{Core selection}

Sediment cores suitable for an $^{60}$Fe search should contain the maximum possible $^{60}$Fe/Fe ratio. Assuming a homogeneous $^{60}$Fe flux over the Earth’s surface, this corresponds to a minimum of terrestrial Fe inputs. Because such inputs originate mainly from continents in the forms of river discharge and dust \cite{Rai11}, pelagic sediments with a continuous sedimentation record over the time of interest (e.g. $1.7-2.7$~Ma) represent the preferred source. Furthermore, selected cores must not have been subjected to major Fe mobilization events, e.g. by reductive diagenesis \cite{Row09}, because such events would redistribute Fe over a much larger depth range, causing (1) $^{60}$Fe dilution, possibly below the detection limit, and (2) incorrect deductions about the event dating and timing. Magnetic iron minerals, and in particular the most strongly magnetic one, magnetite (Fe$_3$O$_4$), are ideally-suited for a systematic characterization of whole core sections in order to check the above-mentioned requirements. In particular, a distinction can be made between coarse primary magnetite crystals originating from rock erosion on one hand, and fine secondary magnetite crystals forming in the water column and the topmost sediment layers on the other hand \cite[e.g. refs.][]{Wat03,Fra07,Yam09}. The latter crystals grow in aqueous solution, inorganically and/or by bacterial mediation, from easily mobilizable Fe sources \cite{Kon11,Tay11,Can88,For05} which include $^{60}$Fe.

The magnetic properties of magnetite crystals depend on their  magnetic domain configuration, which is mainly controlled by the crystal size with the following approximate ranges at room temperature: (1) $<~20~$nm unstable (superparamagnetic) single-domain (SD), (2) $20-100$~nm stable SD, (3) $0.1-1~\mu$m pseudo-SD, and (4) $>~1~\mu$m multidomain \cite{But75,Dun97}. Because the grain size distribution of secondary magnetite extends almost exclusively over ranges (1) and (2) \cite{Lov87,Mah88,Dev98,Fai04} the origin of sedimentary magnetite can be verified with magnetic measurements. Such measurements have allowed us to characterize and quantify the secondary and biogenic magnetite of our sediment cores. A precise, but time-consuming characterization method is described in Sec.~2 below.

A much faster technique suitable for characterizing the magnetic mineralogy of whole core sections consists in imparting a so-called anhysteretic remanent magnetization (ARM) in the laboratory, which is very selective towards stable-single domain crystals \cite{Egl02} and normalizing it with a so-called isothermal remanent magnetization (IRM), which magnetizes all grain size ranges except those of previously mentioned category (1) \cite[e.g. refs.][]{Yam09,Sno02}. ARM is acquired in a slowly decaying alternating magnetic field superimposed to a small constant bias field. We used alternating fields with an initial amplitude of 0.1~T and a bias field $h_{\text{DC}}~=~0.1~$mT, which correspond to typical settings reported in the literature \cite{Egl02}. On the other hand, IRM was acquired applying a field of 0.1~T or 1~T using an electromagnet. ARM and IRM imparted to dried sediment powder pressed into plastic containers have been measured with a 2G Enterprises 755~SRM superconducting rock magnetometer at the paleomagnetic laboratory \cite{Wac12} of the Ludwig Maximilians University (Munich, Germany). This magnetometer has a sensitivity of 10~pAm$^2$, which is fully sufficient for measuring the magnetic moment ($>50$~nAm$^2$) of the imparted magnetizations.

Because ARM intensity is proportional to $h_{\text{DC}}$, the ARM/IRM ratio is usually expressed through the so-called ARM susceptibility $\chi_{\text{ARM}}$, which is the ARM divided by $h_{\text{DC}}$. Accordingly, $\chi_{\text{ARM}}$/IRM is a grain-size sensitive parameter with values $\geq$~3~mA/m characteristic for well-dispersed, inorganic SD magnetite crystals \cite{Egl02} and magnetofossils \cite{Mos93}. Magnetofossil-bearing sediments are systematically characterized by $\chi_{\text{ARM}}$/IRM~>~1 mA/m \cite{Egl10}. ODP cores 848 and 851 considered in this work fulfill this condition over the whole age range under consideration $1.7-2.7$~Ma \cite{Bis11} (see also Fig.~S1).

\section{Magnetofossil detection and quantification}

Reliable magnetofossil detection requires a combination of at least two techniques: (1) TEM observation of magnetic extracts, which serves as a proof for the presence of magnetite crystals with the proper morphology and chemical purity, and (2) magnetic characterization of the bulk sediment for a semi-quantitative assessment of magnetofossil abundance. For better quantification of secondary Fe sources and in particular magnetofossils, we also relied on selective chemical extraction as explained in the following.

\subsection*{TEM observations}

Because of the extremely low concentration of magnetofossils in the sediment samples ($15-35~\mu$g/g), their observation with the electron microscope is only possible after proper magnetic extraction. The magnetic extraction apparatus was built according to the design of ref. \cite{Dob87}, which is in turn similar to other devices used for the same purpose \cite{Hou96}. The extraction procedure starts with the dispersion of $\sim$~7~g sediment in 2~L of distilled and deionised water with an ultrasonic rod. After this step, the sediment suspension is constantly circulated by a peristaltic pump through a glass extraction vessel with a water-tight opening in which a magnetic finger, covered by a teflon sleeve, is inserted. The strong magnetic field gradient in the proximity of the fingertip attracts suspended magnetic particles so that they adhere to the teflon surface. After 24 hours of continuous circulation of the sediment suspension, the magnetic finger is removed and the collected material ($1-2$~mg) is easily harvested once the teflon sleeve is removed from the magnetic fingertip. The iron concentration in the magnetic extract is low (few \%), due to electrostatic adhesion of magnetite crystals to sediment particles \cite[e.g ref.][]{Gal09} and incomplete extraction, but sufficient for electron microscope observations.

Scanning electron microscope analyses (SEM), including energy-dispersive X-ray spectroscopy (EDX) and transmission electron microscope (TEM) analyses, were performed at the Chemistry Department of the Technical University Munich (Germany) for three samples of core 851 in the $2.41-2.62$~Ma age interval where $^{60}$Fe has been found. SEM images (Fig.~S2) reveal the presence of large (>~5~$\mu$m) grains of CaCO$_3$ and SiO$_2$, as well as diatom skeletons, which reflect the typical composition of these pelagic carbonates \cite{Bal95,Gur95}. Large magnetic minerals of lithogenic origin (e.g. titanomagnetite) can also be recognized. TEM images, on the other hand, show abundant iron oxide crystals whose size and shape match those of equidimensional, prismatic, and tooth-shaped magnetosomes (Fig.~S3) \cite{Kop08,Dev98,Man87,Ara05}. EDX mapping of Fe and O matches with these crystals, compatible with magnetite (Fe$_3$O$_4$) composition. Few fragments of the original chain structure can be recognized; however, the original structures are not likely to be preserved by the magnetic extraction procedure, as seen from the comparison of magnetic signatures of extracts \cite{Che07} with respect to that of chemically extractable single-domain magnetite as explained below. The observed magnetofossil crystals are extremely well-preserved, completely lacking corrosion signs indicative of incipient reductive diagenesis \cite{Val87}.

\subsection*{Detailed magnetic characterization}

Electron microscopy of magnetic extracts does not enable quantitative assessments of magnetofossil concentration, especially in relation to lithogenic minerals; nor is it possible to obtain information about the structural integrity of chains, due to the necessity of a possibly disruptive magnetic extraction for sample preparation. Therefore, our electron microscopy observations have been integrated with the most advanced magnetic characterization techniques based on the measurement of partial magnetic hysteresis, in the form of so-called first-order reversal curves (FORC) \cite{Hej90,Pik99}. These curves sweep the whole area enclosed by the major hysteresis loop, thereby enabling a systematic investigation of irreversible magnetic processes, which can be represented as a two-dimensional function \cite{May86} called a FORC distribution. In the context of the Preisach theory \cite{Pre35}, each value of the FORC distribution $f(H_{\text{c}},H_{\text{b}})$ represents the contribution of an elemental rectangular hysteresis loop (hysteron) with coercivity $H_{\text{c}}$ and horizontal bias field $H_{\text{b}}$. In case of uniaxial SD particles, hysterons approximate the hysteresis loops of individual crystals \cite{Nee58}. Hysterons no longer have a physical meaning in case of other magnetic systems; however, natural particle assemblages have typical FORC signatures which reflect their domain state \cite{Rob00}, and, to a certain extent, mineralogical composition \cite{Rob06}.

The advantage of FORC distributions over other magnetic measurements resides in the fact that the signatures of specific magnetic mineral components remain recognizable also in case of complex mixtures \cite{Mux05}. This is particularly true for SD magnetite particles and magnetosome chains that are well dispersed in a sediment matrix. In this case, their signature consists of a sharp horizontal ridge at $H_{b}=0$  \cite{Egl10,New05}, which can be separated from other contributions using appropriate numerical procedures \cite{Egl13}. The separated central ridge is a pure coercivity distribution from which the total saturation magnetization  of SD particles contributing to it can be calculated \cite{Egl10}, obtaining a direct estimate of their mass concentration in sediment. This technique has become a standard method for magnetofossil detection, revealing their widespread occurrence in marine and freshwater sediments \cite{Rob12,Hes13,Hes14}. A detailed FORC investigation of sediment material from the $1.7-3.8$~Ma age interval of core 848 has been reported in ref. \cite{Lud13}. Here, the FORC analysis protocol of ref. \cite{Egl13} has been applied to the untreated sediment material, and to the same material after selective dissolution of ultrafine magnetite particles with a citrate-bicarbonate-dithionite (CBD) solution. Because this treatment removes most secondary Fe minerals while leaving large lithogenic crystals intact \cite{Hun95,Vid00}, it provides an independent manner for identifying magnetofossils and other secondary magnetite crystals, thereby excluding possible contributions from SD magnetite particles enclosed in primary silicate minerals, which would not contain $^{60}$Fe. A summary of FORC analysis results from ref. \cite{Lud13} is provided with Fig.~S4. The main outcomes of such analysis are summarized in the following:
\begin{enumerate}
\item The sediment contains SD particles and chains of such particles, which are directly dispersed in the sediment matrix and not hosted inside lithogenic minerals. This proves their secondary origin as inorganic or biologically mediated precipitate, and as magnetofossils. The presence of magnetofossils is confirmed by TEM observations, while few irregularly shaped magnetite crystals compatible with a non-magnetofossil origin could be found. The saturation magnetization of these particles, as deduced from the central ridge of the FORC function, corresponds to a Fe mass concentration of $2.7\times10^{-5}$, which represents at least 27\% of our CBD-extractable iron. Details about the calculation of Fe mass concentrations from SD magnetite particles are reported in ref. \cite{Fai08}.
%
%
\item The difference between FORC diagrams measured before and after CBD treatment reveals a second contribution to the FORC function, which is also attributable to ultrafine (SD) magnetic particles. This contribution bears the signature of magnetostatic interactions and might be attributed to particle clusters, whose origin is not clear. They might be the product of chemical precipitation, as well as the result of magnetofossil chain collapse \cite{Kob06}. Nevertheless, their secondary origin is supported by the fact that such particles are not located inside lithogenic minerals (e.g., silicates), and that application of the same chemical treatment to volcanic ash, which represents one of the major lithogenic sources of equatorial Pacific sediment \cite[e.g. ref.][]{Str98} did not remove a significant amount of magnetic minerals. The saturation magnetization of these particles corresponds to a Fe mass concentration of $1.5\times10^{-5}$, i.e. 15\% of our CBD-extractable iron.
\item From the above-mentioned estimates, the total mass concentration of Fe from secondary magnetite crystals is $4.2\times10^{-5}$, totaling 42\% of CBD-extractable iron. This also represents an upper limit for the contribution of magnetofossils.
\item The FORC signature of sedimentary greigite (Fe$_3$S$_4$) \cite{Rob12}, an iron sulphide produced exclusively during reductive diagenesis \cite{Fu08}, is completely absent from the investigated sediment. This is also confirmed by the lack of magnetosomes with signs of corrosion in TEM images of magnetic extracts. Magnetofossil preservation implies that no major Fe transport or diagenesis occurred after the formation of magnetosomes, i.e. after their permanence in the so-called benthic mixed layer (BML), i.e. the topmost layer of the sediment column that is permanently mixed by benthic organisms.
\item Assuming $L\approx$~7~cm for the typical thickness of the surface mixed layer in pelagic carbonates, along with an estimated sedimentation rate $r_\text{sed}\approx0.61$~cm/ka for core 848 over the $1.7-2.7$~Ma age interval (Fig.~1 and ref.~\cite{Pis95}), we obtain a mean residence time $t_{\text{BML}}\approx 13~$ka for magnetofossils in the BML. Variations in $L$ and/or $r_\text{sed}$ over time produce an age uncertainty related to the change of $t_{\text{BML}}$. If maximum variations in total Fe concentration and $\chi_{\text{ARM}}$/IRM over the time range of interest, which correspond to a factor $\sim$~2, are attributed to environmental changes that have a similar effect on $t_{\text{BML}}$, the maximum age uncertainty of Fe in secondary magnetite can be set to 26~ka, i.e., a negligible fraction of the total duration of the $^{60}$Fe signal ($1.0 \pm 0.3$~Ma).
\end{enumerate}

\section{$^{60}$Fe AMS sample preparation}

The main goal of the chemical extraction and purification procedure was the production of AMS samples with a $^{60}$Fe/Fe ratio being as close as possible to the ratio in the targeted minerals, namely secondary iron oxides such as magnetofossils. This requires, on the one hand, that the dilution with stable Fe from other sources is minimized, and that no contaminations with $^{60}$Fe from other sources can occur. The following procedure has been shown to accomplish these goals remarkably well, as described in ref.~\cite{Lud13}, where a more detailed description can be found.

For the preparation of the CBD procedure, $\sim$~35~g of dry sediment were crushed in an agate mortar and then added to 200~ml of distilled and deionised H$_2$O. Under constant stirring, the temperature was brought to ($50\pm2$)$^{\circ}$C on a hot plate while adding 3.4~g of sodium bicarbonate and 12.6~g of sodium citrate. The reaction is then started by adding 5.0~g of sodium dithionite, a strong reducing agent. This corresponds to the concentrations suggested by ref. \cite{Hun95}. The main extraction mechanism relies on Fe(III) reduction to Fe(II) by sodium dithionite on mineral surfaces. Fe(II) is then chelated by sodium citrate, while the pH is kept stable at 7.3 by the sodium bicarbonate.

The procedure was fine-tuned to selectively dissolve $<~200$~nm Fe$_3$O$_4$ crystals. This ensures that secondary iron oxides are completely dissolved, while primary ones are left essentially intact. To this end, an extraction temperature of 50$^{\circ}$C and an extraction time of 1~h were chosen.

After the extraction, the remaining undissolved sediment material was separated from the Fe(II) solution using a filter paper of 0.1~$\mu$m pore size. The Fe(II) solution (clear yellow in color due to the presence of dissolved Fe(II)) was then evaporated 
to dryness. In order to destroy the sodium citrate chelation, the sample was then heated for 1~h at 300$^{\circ}$C, decomposing the citrate, until a black, carbon-rich solid was obtained. Fe(II) was then oxidized by adding 50 mL HNO$_3$ (65\%). After another heating step for 1 h at 400$^{\circ}$C, all carbon was oxidized and the sample became a colorless liquid that solidified upon cooling to room temperature. Fe(III) was then extracted using 30~mL HCl (7.1~M), which was evaporated to dryness. Another 20~mL HCl were added and much of the organic residuals could be removed by centrifugation. This step was repeated by dissolving in another 20~mL HCl, evaporating again, adding another 20~mL HCl and centrifugation. From the final HCl solution containing Fe(III), iron hydroxide was precipitated by adding NH$_3$(aq) (25\%) was possible. The precipitate was washed 3 times with slightly alkaline solution (pH~8-9) and then re-dissolved in 1.5~mL HCl (10.2~M). In order to remove unwanted contaminations by other metals that might have precipitated along, an anion exchange (DOWEX 1x8) step followed. This was done similarly to the procedure described in ref. \cite{Mer99}. After re-precipitating the iron hydroxide and another 3 washing steps with deionized water, the samples were transferred to quartz crucibles and baked at 600$^{\circ}$C for 3 hours, resulting in samples of Fe$_2$O$_3$ with a typical mass of 3-5~mg, sufficient to fill one AMS sample holder.

The recovery efficiency of this procedure was determined using commercial magnetite grains (40-60~nm, Alfa Aesar, Lot Nr. E08T027) to be ($85\pm5$)\%. The contamination with stable Fe from the chemicals was found to be less than 0.05~mg per sample. 

\section{$^{60}$Fe measurements at the GAMS setup at MLL}

\subsection*{Ion source}
The ion source used for the experiments described here is a home-made \cite{Sch07}, single cathode, Middleton type \cite{Mid83} cesium sputter ion source with a spherical ionizer \cite{Urb86} optimized for optimum mass resolution. The Cs vapor is ionized on the tantalum ionizer and accelerated towards the sample using a 5~kV sputtering voltage. The negative ion beam is then extracted by applying an additional voltage of 23~kV. After preliminary mass separation by a dipole magnet, a beam of $^{54}$FeO$^-$ is selected and directed towards the accelerator entrance with several electrostatic lenses. After a further pre-acceleration to 178~keV, the beam is injected into the accelerator, at which point the typical current of $^{54}$FeO$^-$ is 30–150~enA.

\subsection*{Ionization chamber} 
The particle identification detector is an ionization chamber filled with 40–60~mbar of isobutane, featuring a 5-fold split anode. The first two anodes sections are additionally split diagonally, to provide the x-position and angle of the incoming particles by comparing the left and right energy loss signals each. In total, particle discrimination is possible with the 5 $\Delta$E anode signals, the Frisch grid signal (proportional to the sum of all $\Delta$E signals), the x- and y-angle of the incident particles, and their x-position. The discrimination between the detector signals, from each anode, arising from the passage of ions of different atomic charge is possible on the basis of differential ionization energy loss (Bragg curve spectroscopy) of the ions passing through the detector gas. This also provides additional discrimination between the isobars $^{60}$Fe and $^{60}$Ni since they differ in their proton number, but also between different isotopes of the same element and other background.

\subsection*{Standard sample} 
For all measurements discussed in this work, $^{60}$Fe was measured relative to the standard sample PSI-12. The starting point for the production of this standard was material extracted from a beam-dump at the Paul Scherrer Institute (Switzerland), which was later used for the half-life measurement of $^{60}$Fe in ref. \cite{Rug09}. The high concentration of $^{60}$Fe/Fe was precisely measured by ICP-MS. This material was diluted, by stable Fe, to a ratio of $^{60}$Fe/Fe$~=~(1.25~\pm~0.06)\times~10^{-12}$ and referred to as PSI-12 for clarity. In order to avoid cross-talk in the ion source, during each standard run (about every 3 hours), only about 200 events of $^{60}$Fe are recorded, since the exposure of the ion source to a highly concentrated sample (in this case 4-5 orders of magnitude above the blank level) must be minimized. The possibility of cross-talk was examined by first sputtering a sample with a concentration of $^{60}$Fe/Fe~$\approx~10^{-9}$ for 20 minutes (typical time for a standard measurement) and subsequently measuring a blank. This way, cross-talk from one sample to the next could be determined to be $\sim 10^{-6}$.

\subsection*{$^{60}$Fe event selection}
The $^{60}$Fe events from the standard sample PSI-12 allow for empirically constructing, for each anode, their respective distribution/histogram of collected ionization charge from the transit of $^{60}$Fe in the detector. Knowing these distributions/histograms then allows for the identification of $^{60}$Fe from the actual sediment samples, of much lower $^{60}$Fe concentration. This procedure is illustrated in Fig.~S5. In order to reduce the 6-dimensional space of the energy signals of an event, it is useful to introduce the $\chi^2$-value determined as the sum of squared deviations of an event from the standard sample position of $^{60}$Fe normalized to the width of each respective peak, i.e.

\begin{equation}
\chi^2=\sum_{i=0}^{5}\frac{(E_i-\mu_i)^2}{\sigma_i^2}
\end{equation}

Here, $E_i$  represent the 6 energy signals from the detector (with  corresponding to the Frisch-grid signal) for the candidate event of $^{60}$Fe under consideration, and $\mu_i$  and $\sigma_i$ are the position and width of the $^{60}$Fe event distribution obtained from the standard sample, respectively. The maximum of the distribution for a standard sample is not (as would be expected for a $\chi^2$ distribution with 6 degrees of freedom) about 4, but rather around 2.5, as shown in Fig.~S6(A), which can be explained by taking into account that the energy signals are not completely independent and that for some parameters the tails of the distributions are not in the accepted signal range. Nonetheless, the $\chi^2$-value of a candidate event has high predictive power, i.e. a low $\chi^2$ favors a real $^{60}$Fe event. In most cases of $^{60}$Fe measurements, a 1-dimensional cut on the $\chi^2$ (accepting only events with $\chi^2~<~15$), combined with a 2-dimensional cut on the X-position and one of the energy signals (choosing the one with best separation - typically $\Delta$E3) was sufficient for event discrimination. The $\chi^2$ distributions obtained from the $^{60}$Fe events from both the standard sample and all sediment samples can be seen in Fig.~S6(B). Both distributions agree extremely well, confirming the authenticity of the observed $^{60}$Fe signal.

\subsection*{Discussion of AMS uncertainties}
All AMS results in this work are presented with 1-$\sigma$ confidence intervals. Owing to the relative measurements to a standard sample, all systematic uncertainties related to transmissions and efficiencies cancel. The only systematic uncertainty that cannot be avoided is the uncertainty of the concentration of the standard sample itself (4.8\%). This would however only shift the entire AMS $^{60}$Fe/Fe data up or down and was thus omitted.

The statistical errors that were included in all AMS results have several origins. The low count-rate of $^{60}$Fe events is the main source of uncertainty. This was treated by employing the confidence intervals suggested by ref.~\cite{Fel98} for zero background. The uncertainty in the ion current reading was estimated to be 15\% in each data run. Additionally, the fluctuation of the atom ratio $^{60}$Fe/Fe measured in the standard sample was typically 15\%. Another 20\% uncertainty is added due to the unknown behavior of the ion current during runs. With an average of 4 data runs per sample, the quadratic summation yields a statistical uncertainty of 15\% for each sample, which is a rather conservative estimate. For each data point (e.g. grouped samples in Fig.~3), this number is divided by the square-root of the number of samples used to produce the grouped data point. This is then added in quadrature to the relative uncertainty of the confidence intervals given by ref.~\cite{Fel98} , except in the case of zero events, where the 1-$\sigma$ upper limit of 1.29 events is conservatively increased in all cases by 0.15 (corresponding to 15\% at 1 event) to yield 1.44 events.

The horizontal error bar associated with each data point in the $^{60}$Fe/Fe plots represents the time-span interval over which raw sediment material was used in our CBD protocol in order to obtain a sufficiently large mass of Fe$_2$O$_3$ extract suitable for an AMS sample (at least 3~mg). The x-position of each data point is the average x-position of individual samples used for the grouping, weighted by the amount of statistics collected for each sample (i.e. the product of average beam current and measuring time). The systematic uncertainty in core dating is not included in these error bars. One possible such data grouping is shown in Fig.~4, after correction for radioactive decay of $^{60}$Fe.

\section{Temporal structure of $^{60}$Fe signal}
In order to extract the temporal structure of the $^{60}$Fe input into the sediment samples, the obtained $^{60}$Fe/Fe AMS data has to be corrected for the blank level and the radioactive decay of $^{60}$Fe. Using the established constant sedimentation rates (Fig.~\ref{fig:Fig1}), this ratio is directly proportional to the deposition rate of $^{60}$Fe atoms, as shown in Tab.~1 for representative time intervals. The deposition rates corresponding to the data grouping of Fig.~\ref{fig:Fig2} are shown in Fig.~S7 and take into account variations in the CBD-extracted Fe yield of every AMS sample.

\section{Sample preparation and $^{10}$Be measurements at DREAMS}

To further independently establish the constancy of the sedimentation rate, 13 sediment samples of core 851, in the depth (age) range 1.7~Ma to 3.0~Ma, were analyzed for $^{10}$Be ($t_{1/2}=1.38$~Ma, refs. \cite{Chm10,Kor10}). $^{10}$Be is mainly produced in atmospheric spallation reactions and can be transported to Earth’s surface by dust, rain, and snow. Upon its deposition into geological reservoirs such as ice cores and marine sediments, it can be used for dating over several Ma \cite{Bou89}.

From each depth, $\sim$~3~g of sediment were subjected to a hydroxylamine-based leaching technique, described in ref. \cite{Fei13}. The AMS measurements of $^{10}$Be were performed at the DREAMS facility at the Helmholtz-Zentrum Dresden-Rossendorf, using the procedure described in ref. \cite{Akh13}. The determined concentrations of $^{10}$Be/$^{9}$Be follow the general trend of the exponential decay, which is expected for an undisturbed core of constant sedimentation rate. This is shown in Fig.~S8. The exponential fit to the data using the known $^{10}\mbox{Be}$ half-life ($t_{1/2}=1.387$~Ma) produces a reduced $\chi^2$ of 1.25 and no significant deviation from an exponential decay as a function of age are evident, demonstrating again that our cores are characterized by constant sedimentation rates across the time window spanning 1.7~Ma to 3.0~Ma, consistent with the magnetostratigraphic models of ref. \cite{Sha95}. 

\section*{Supplementary Figure captions}

\begin{itemize}
\item \textbf{Figure S1} $\chi_\text{ARM}$/IRM vs. sediment age for core 851 composed from two different measurements. The systematic offset between the blue and the red curve is due to the maximum field used to impart the IRM, which was 0.1~T for the red curve and 1~T for the blue curve. IRM acquired in the larger field includes the contribution of high-coercivity minerals which are not magnetized during the measurement of $\chi_\text{ARM}$, therefore yielding lower values of $\chi_\text{ARM}$/IRM. Overall, $\chi_\text{ARM}$/IRM displays minor variations and is consistently compatible with values expected for magnetofossil-rich sediments.

\item \textbf{Figure S2} Scanning electron microscopy images of magnetic extracts obtained from representative sediment samples of core 848 over the 2.41–2.62 Ma age interval. (A) Overview of the extract showing prominent features including (1) the copper sample holding grid, (2) large grains consisting mainly of CaCO$_3$ and SiO$_2$, and (3) diatoms. (B) High-resolution image of a titanomagnetite grain (dark-gray octahedron in the center) of most likely lithogenic origin. (c) High-resolution image of a silicate grain (dark, in the center) with small-grained Fe-bearing minerals adhering on its surface (bright spots). All images have been obtained with a JSM5900V SEM (Jeol, Japan).

\item \textbf{Figure S3} Transmission electron microscopy images of magnetic extracts obtained from representative sediment samples of core 848 over the 2.41–2.62 Ma age interval showing abundant magnetofossils. All images have been obtained with a JEM2011 (Jeol, Japan).

\item \textbf{Figure S4} Magnetic analyses of a representative sample from core 841. (A) FORC measurements of the untreated sediment (every 8th curve shown for clarity). (B) Same as (A) for the residue of CBD extraction. (C) FORC diagram calculated from measurements in (A). (D) FORC diagram calculated from measurements in (B). Notice the $\sim 20$ times smaller amplitudes in comparison to (C). (E) Difference between C and D, which can be identified with the magnetic signature of CBD-extractable particles. (F) Coercivity distributions associated with the central ridges in (C-E) (i.e. the ridges along $H_b=0$). Almost all SD particles responsible for these ridges can be extracted with the CBD procedure. (G) Same as (F), where the central ridge coercivity distribution of the CBD residue has been multiplied by 10 in order to show the fundamentally different distribution shape with respect to that of extractable particles.

\item \textbf{Figure S5} Illustration of the event selection process for $^{60}$Fe data analysis with representative data from sediment core 851 – 2.35-2.36 Ma. Energy loss in anode 3 and X-position spectra from a standard sample (A), a blank sample (B), the representative sediment sample (C) are compared. Spectra (D-F) are produced by applying 1-dimensional cuts on all energy signals except $\Delta$E3 and the incident X- and Y-angles. If necessary, a final 2-dimensional cut is applied to one of the energy-loss signals (in this case $\Delta$E3) and the X-position, as indicated in (D). This results in a final, 9-dimensional region of interest for $^{60}$Fe events, which can be applied to the blank (E) and the sediment sample (F), yielding the number of events and thus the concentration of $^{60}$Fe/Fe.

\item \textbf{Figure S6} (A) The $\chi^2$ distribution of all ($\sim 30.000$) $^{60}$Fe events collected from standard samples over all beamtimes (black histogram) is fitted with a $\chi^2$-density function of variable amplitude and number of degrees of freedom (NDF). The best fit is produced using NDF = 4.5 (red line). The fit has a $\chi^2$/NDF-value of 1.7. (B) A comparison between the $\chi^2$ distribution of all 89 $^{60}$Fe events in the sediment samples with the expected distribution obtained from the events in the standard sample (scaled from (A)). The last bin of (B) sums up all events in the range $7\leq \chi^2 \leq 15$.

\item \textbf{Figure S7} Flux of $^{60}$Fe into the sediment cores deduced from the decay and blank corrected $^{60}$Fe/Fe ratios determined in all sediment samples of core 848 (A) and 851 (B). The data use constant sedimentation rate and assume constant influx of CBD extractable Fe. Y-error bars indicate 1-$\sigma$ statistical uncertainties. X-error bars represent core depth of sample material used for data point. Each data point contains 3-6 adjacent individual samples grouped together. These data are representative of the temporal structure of the SN $^{60}$Fe signal. Note that panel  (A) has a broken time axis relative to panel (B).

\item \textbf{Figure S8} AMS results of $^{10}$Be from core 851 sediment samples. Horizontal error bars (almost negligible) represent the sampling range and vertical error bars correspond to 1-$\sigma$ confidence intervals. The red line is a least-squares ($\chi ^2$/NDF~=~1.25) fit obtained using a fixed $^{10}$Be half-life of 1.387~Ma.
\end{itemize}




\end{document}